\documentstyle[]{l-aa}
\input psfig.tex
\begin{document}
\topmargin 25mm 
\thesaurus{02.13.3; 09.07.1; 09.13.2; 11.09.4; 11.19.3; 13.19.2 }

\title{Search for 183 GHz water maser emission in starburst galaxies }

\author{F. Combes, Nguyen-Q-Rieu, Dinh-V-Trung }

\institute{ DEMIRM, Observatoire de Paris,
 61 Av. de l'Observatoire, 75\ts014 Paris, France}

\offprints{F.~Combes (e-mail: bottaro@obspm.fr)}
\date{Received Sep 1996; accepted Jan 1997}
\maketitle
\markboth{Combes et al.:
183 GHz water maser emission in starburst galaxies }{}

\begin{abstract}
We have searched for water vapor emission at 183 GHz, redshifted at 157 GHz
and 161 GHz, in the two ultraluminous starburst galaxies Mrk1014 and
VIIZw244. Due to the low energy level of the upper state of the 
183 GHz transition ($\approx$ 200K), it is expected that the emission regions 
are extended, as they are in the Orion or W49N molecular cloud cores. 
Since the warm and dense gas, typical of star-forming cores, 
is expected to have a large
surface filling factor in ultraluminous starburst galaxies, the
maser H$_2$O emission at 183 GHz could have been detected.
In fact, no water line has 
been detected in Mrk1014 and VIIZw244 with an upper limit of $\sigma$ = 1 mK.
We compare the H$_2$O/CO emission ratio with that obtained towards
the ultraluminous high-z object F10214+4724, which flux is amplified by
a gravitationnal lens. We suggest that the
amplification factor for the H$_2$O emitting cores in this galaxy
should be higher than for the general CO emitting region.
We conclude that the warm and dense H$_2$O cores are much less extended
in the two observed starburst galaxies than in the Orion molecular cloud;
this provides some information on the physical conditions and 
cooling processes of the interstellar medium in starburst galaxies.

\end{abstract}

\section{ Introduction } 

 Water is thought to be one of the most abundant molecule of the interstellar
medium. Unfortunately, the broad atmospheric lines prevent any direct detection
in our own Galaxy. Attempts have been made, through the isotopic molecules
HDO and H$_2^{18}$O (Henkel et al 1987, Jacq et al 1988, 1990, Wannier et al
1991, Gensheimer et al 1996), and through the precursor ion H$_3$O$^+$ 
(Phillips et al 1992), and abundances of the main isotope of H$_2$O around 
10$^{-5}$ have been deduced. This is also confirmed by the detection in Orion 
of absorption lines at 2.66 microns with the K.A.O. (Knacke \& Larson 1991).
Very recently, observations with the ISO satellite of the $2_{12}$ -- $1_{10}$
179.5 $\mu$m line of ortho water in absorption against the continuum of 
the galactic center (SgrB2, Cernicharo et al 1997) has revealed that the
H$_2$O molecule is abundant over very extended regions. 

   Water levels can be collisionally pumped in the dense (n(H$_2$) =
10$^8$-10$^{10}$ cm$^{-3}$) star-forming 
regions to emit strong maser lines: the 
6$_{16}$-5$_{23}$ line at 22GHz has been widely observed, because of the 
atmospheric transparency at this wavelength, and the 3$_{13}$-2$_{20}$
line at 183 GHz has been detected towards the Orion molecular cloud
with the K.A.O. (Waters et al 1980), as well
as the 4$_{14}$-3$_{21}$ line at 380GHz (Phillips et al 1980). All these
lines are well known to be masing, since they correspond to "backbones",
i.e. levels with the lowest energy for a given rotational quantum 
number (de Jong 1973, Cooke \& Elitzur 1985, Deguchi \& Nguyen-Q-Rieu 1990).  
Neufeld and Melnick (1987) computed the total
flux emitted by the H$_2$O lines in the shocked gas region of Orion. They 
concluded that H$_2$O is by far the main coolant, 40 \% of the 
total cooling being provided by the far-infrared ortho-H$_2$O lines, 
in the shocked gas regions.
Since then a series of higher-level maser lines have been detected
(Menten et al 1990a,b; Melnick et al 1993). Neufeld and Melnick (1990, 1991)
interprete the maser data in the frame of excitation models, and show
that the minimum masing gas temperature is 200K. This favors slow 
($\la$ 50 km/s) non-dissociative shocks for the source of maser emission
(Melnick et al 1993, Kaufman \& Neufeld 1996a,b). 
 
Even from the ground at 183 GHz, Cernicharo et al (1990) succeeded,
with the IRAM 30m telescope, to detect in Orion the para H$_2$O line which
turns out to be inverted with a huge flux of a few 10$^4$ Jy (main beam 
brightness temperatures of the order of 2000K).
Contrary to the point-like maser sources at 22 GHz, the 183 GHz
maser emission is quite extended ($\sim$ 1 arcmin = 0.2pc), and represents a 
significant coolant in the star-forming region (Cernicharo et al 1994).
Gonz\'alez-Alfonso et al (1995) also found that the 183 GHz emission is
spatially extended in the star-forming cloud W49N. The maser dominated
183 GHz H$_2$O emission is
not confined to narrow shock regions, but on the contrary extends out to
2.2pc from the cloud center, and is very similar in extension and kinematics
to molecular emission from dense gas tracers (CS, HCN, H$_2$CO..).
Also extended 183 GHz H$_2$O maser emission has been detected in low-mass
star forming regions, such as HH7-11 (Cernicharo et al 1996).
Cernicharo et al (1994) show, using an escape probability model
for radiative transfer, that the 183 GHz line can be inverted for densities
n(H$_2$) $\sim 10^5$- 10$^6$ cm$^{-3}$ and kinetic temperatures between
50 and 100K. The maser begins to saturate for a para-H$_2$O column density
larger than 10$^{18}$ cm$^{-2}$. Modelisation of the Orion region 
implies that the H$_2$O/CO abundance ratio is 0.3-0.5.

The detection of the 183 GHz transition of water
should provide additional constraints to investigate the physical conditions
in the extended star-forming regions that are found in the central region
of starburst galaxies.   
  For this purpose, we have tried to detect the masing 183 GHz line in two 
infrared ultraluminous
starburst galaxies, redshifted in a transparent window of the atmosphere, near 
$\lambda$=2mm. We have chosen two mergers clearly detected in CO 
 at a velocity near 50 000 km/s with a redshift
ideal for our purposes: Mrk 1014 (Sanders et al 1988b, Alloin et al 1992)
and VIIZw244 (Alloin et al 1992). No H$_2$O line
was detected in both galaxies with a 3$\sigma$ upper limit
of 3 mK. We discuss the main implications of these negative results on the 
determination of the physical parameters
of the interstellar medium in starburst galaxies.

\section { Observations }

The observations were made with
the IRAM 30m telescope at Pico Veleta near Granada, Spain, in August 1993.
Table 1 displays the observational parameters. 

We observed simultaneously with three SIS receivers: at 3mm, we tuned
to the CO(1-0) frequency to check with the previous observations of
Alloin et al. (1992) that the pointing and calibration were optimum  
to obtain a high quality CO spectrum; at 2mm, we observed the redshifted 183 
GHz H$_2$O line, and at 1.3mm we searched for the HCN(3-2) line. We used  
two 512x1MHz filterbanks and an AOS backend. The observing procedure was
wobbler switching with a small throw in azimuth (1'). We calibrated the
temperature scale every 10 minutes by a chopper wheel on an ambient temperature
load, and on liquid nitrogen. Pointing was checked on broadband continuum
observations. The relative pointing offsets between the 3 receivers were
of the order of 3". We checked the frequency tunings by observing molecular
lines towards Orion, DR21 and IRC+10216.

We integrated for 7 and 15 hours on Mrk1014 and VIIZw244 respectively.
We obtained a noise-level of 1 mK at the H$_2$O frequency for both
sources, at 60 and 29km/s resolution. The system temperatures 
were on average 300, 600 and 700K at 3, 2, and 1.3mm respectively.

\begin{table}
\begin{flushleft}
\caption[]{ Observational parameters }
\begin{tabular}{|lcc|}
\hline
 & & \\
\multicolumn{1}{|l}{ }                           &
\multicolumn{1}{c}{Mrk 1014}	              &
\multicolumn{1}{c|}{VIIZw244}                  \\
  & & \\
\hline
 & & \\
$\alpha$(1950)	& 01$^h$ 57$^m$ 15.8$^s$ & 08$^h$ 38$^m$ 32$^s$    \\
$\delta$(1950)	& 00$^\circ$ 09$'$ 10$"$& 77$^\circ$ 03$'$ 59$"$ \\
    z		& 0.1631	& 0.1324	\\
 & & \\
\hline
 & & \\
 I(CO) (Kkm/s)	& 1.0		& 0.76		\\
$\Delta$V$_{CO}$ (km/s)& 210.		& 80.		\\
 T$_{mb}$(CO) (mK)  & 4.5		& 9		\\
 & & \\
\hline
 & & \\
I(H$_2$O) (Kkm/s) & 0.6$^*$ 	& 0.2$^*$	\\
 & & \\
\hline
\end{tabular}
\, \\
\vskip 4truemm
$^*$ 3$\sigma$ upper limit \\
\end{flushleft}
\end{table}

\begin{figure}
\psfig{figure=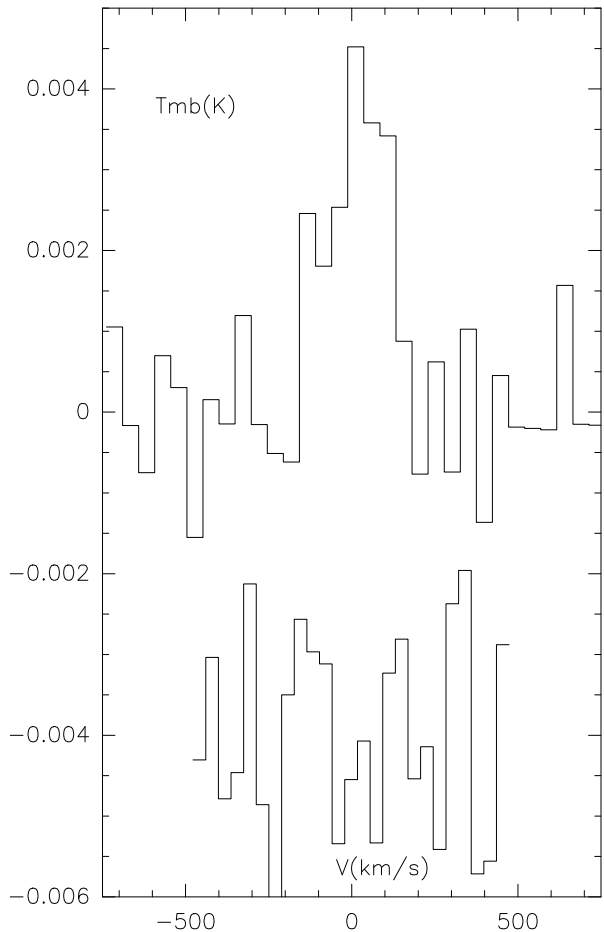,height=13.5cm,angle=0}
\caption[]{ Spectra at the frequencies of CO(1-0) and H$_2$O 183 GHz for the
ULIGs Mrk 1014. The temperature scale is T$_{mb}$ }
\label{maser_fig1}
\end{figure}

\begin{figure}
\psfig{figure=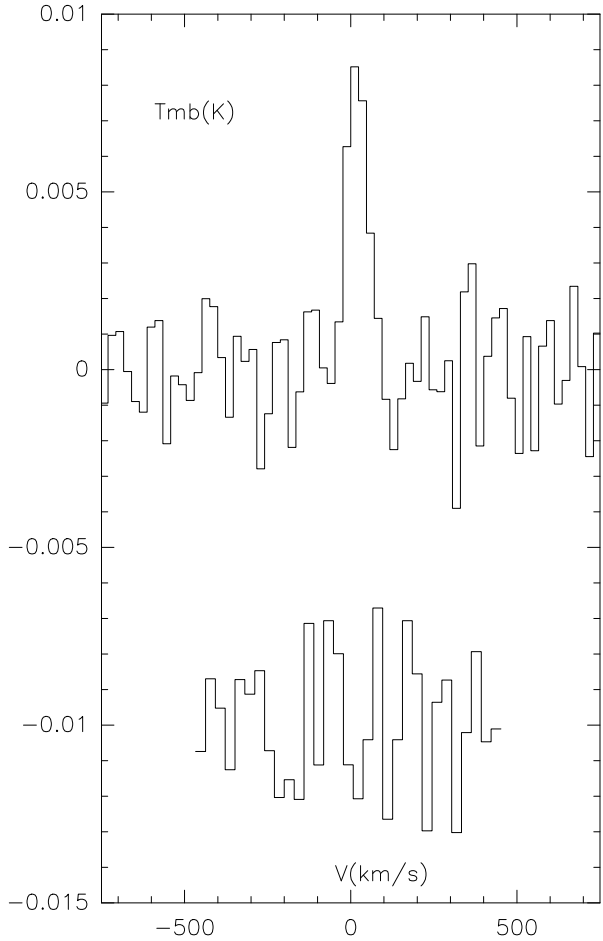,height=13.5cm,angle=0}
\caption[]{ Same as fig. 1 for VIIZw244 }
\label{maser_fig2}
\end{figure}

Figures \ref{maser_fig1} and \ref{maser_fig2} present our CO (1-0) spectra for 
Mrk 1014 and VIIZw244, together with the upper limits for the 183 GHz water 
line redshifted at 157 and 161 GHz, respectively.
 We derived the 3$\sigma$ upper limits in 
Table 1 by assuming the same linewidths in CO and H$_2$O. The temperature
scale is the main beam brightness temperature. The main beam efficiencies are
 0.56 and 0.52 at 3 and 2mm, respectively.

We obtain for Mrk 1014 and for VIIZw244 a CO integrated intensity almost 
consistent with that obtained by Alloin et al (1992), although 10 to 
30\% lower. However, we note that in both cases their spectrum is composed of 
a narrow feature, that we retrieve entirely, and of a wider shoulder
that could be due to the baseline, since some wavy structure is also 
seen beside. If the shoulder is included in the baseline, then our two
spectra are compatible. Their reported CO FWHM are indeed 10 to 30\%
higher than our corresponding values.

No HCN (3-2) was detected, with an upper limit at 3$\sigma$ of 9 and 6
 mK in 10km/s channels, for Mrk1014 and VIIZw244 respectively. This is
perfectly compatible with what is expected from starburst galaxies, i.e.
an HCN(1-0) intensity of the order of 20\% of the CO(1-0) intensity 
(Solomon et al 1992a); HCN(3-2) is even weaker (Nguyen-Q-Rieu et al
1992).

\section { Discussion }

\subsection { Dense cores in ultraluminous starburst galaxies }

The spectacular ultra-luminous IRAS galaxies (ULIGs) have been recognized
to be powered by violent starbursts triggered by tidal interactions and
mergers (e.g. Sanders et al 1988a). They are among the most luminous
objects in the Universe, and radiate at least 90\% of their energy in
the far infrared. Their molecular content is about 20 times that of a 
normal galaxy such as the Milky Way, and their star formation efficiency,
traced by the L$_{FIR}$/M(H$_2$) ratio, is 10 to 50 times larger
than for normal molecular clouds (Solomon \& Sage 1988, Sanders et al 1991).
 Interferometric measurements of CO emission have revealed that the
molecular component is strongly concentrated in the central kpc, and
the average surface density can be as high as 10$^5$ M$_\odot$/pc$^2$,
corresponding to an H$_2$ column density of 5 10$^{24}$ cm$^{-2}$ 
(Scoville 1991). This gas is for a large part at much higher densities
than in normal Giant Molecular Clouds, as revealed by HCN observations
(Solomon et al 1992a): this polar molecule traces gas at density larger
than 10$^5$ cm$^{-3}$, while the CO molecule traces gas at $\le$
500 cm$^{-3}$. All these properties of the molecular component in ULIGs
suggest that most of the gas is in very dense regions similar to star
forming cores, such as the Orion molecular core. The surface filling
factor of these dense and warm cores, which is very small in a normal
galactic disk (about 10$^{-8}$) should be enhanced here by 4 to 5
orders of magnitude, up to $\approx$ 10$^{-3}$. 

The most extreme example of these objects might be the high redshift
IRAS source F10214+4724 (Rowan-Robinson et al 1993), with an apparent 
luminosity 1.8 10$^{14}$ L$_\odot$ (H$_0$ = 75 km/s/Mpc and q$_0$=0.5).
At the redshift of z=2.286, the apparent CO luminosity implies a huge 
H$_2$ mass larger than 10$^{11}$ M$_\odot$ (Solomon et al 1992b).
However there is now evidence that this source is amplified by an
elliptical galaxy in the foreground (Elston et al 1994, Broadhurst \& Leh\'ar
1995, Graham \& Liu 1995). In particular, HST images have shown that the main 
image is an arc, with an amplification of the order of 100 (Eisenhardt et al 
1996). This high amplification factor concerns only the smallest optical
source (of intrinsic size of 80pc), while the FIR amplification 
could be 30, and the molecular
emitting region is amplified by an uncertain factor between 5 and 50.
 If this is taken into account, the F10214+4724 molecular content is
perfectly similar to that of other ULIGs. The only other
high redshift ULIGs detected up to now, the clover-leaf at z=2.556 
(Barvainis et al 1994) and Q1202-0725 at z=4.69 (Omont et al 1996)
are also suspected to be strongly amplified (several components are 
detected); it then appears that ULIGs could be the most extreme objects
as far as molecular emission is concerned, and no intrinsically brighter
object has been detected until now. Moreover, we cannot
detect them at high z if they are not gravitationally amplified (Evans et al 
1996, Solomon et al 1996).

\subsection { Surface filling factor of dense cores }

Given the high average molecular surface density, the high volumic
density deduced for most of the gas, the high star-forming efficiency
and the high dust temperature (between 50 and 100K), 
we can expect that such actively star-forming regions like the Orion dense
core are quite numerous in ultra-luminous galaxies.
In fact, observations show that most of the molecular emission comes
from these warm dense cores (Solomon et al 1996).

 For example, water thermal emission has been tentatively detected at 752 
GHz in the starburst IRAS galaxy F10214+4724
(Encrenaz et al 1993, Casoli et al 1994), with 
a surface filling factor only slightly smaller than that of CO emission. 
This can be found from the comparison of observed
antenna temperatures between CO and H$_2$O spectra obtained with 
the IRAM 30m telescope.
 Since both the CO lines ((3-2), (4-3) and (6-5)) and
the 752 GHz H$_2$O line (2$_{11}$ -- 2$_{02}$ line
of para water) are highly optically thick, 
no abundances or column densities can be deduced directly from observations,
but antenna temperatures give instead an estimate of the 
surface filling factor.
The CO data (Solomon et al 1992b,
Brown \& VandenBout 1992), have shown that the emission cannot
be interpreted in terms of a single component at a given density and
temperature, unless the average density is sufficiently low, such that the CO
excitation is subthermal (the apparent peak brightness
temperatures are in the ratio 1:0.55:0.16 for the CO 3-2:4-3:6-5 lines). 
The emission is more realistically coming from various components of
different temperatures and densities, and it might be more interesting 
here to compare the equivalent surface filling factors of the various
line emissions, as if they came from the same brightness temperature
regions in F10214+4724.
The brightness temperature of the emitting dense gas
should be of the order of the kinetic temperature $\sim$80 K (the 
measured FIR dust temperature). The rest-frame brightness temperature T$_b$ 
has to be redshifted to T$_b$/(1+z)= 24K (for $z=2.28$) to give the apparent
brightness temperature. Then, to get the measured antenna temperature, 
we have to take into account the size of the source at mm wavelength,
of the order of 1.5"x0.9" (Downes et al 1995), and the dilution factor in the
telescope beam (24" at CO(3-2), 18" at CO(4-3), 12" at CO(6-5) and H$_2$O). 
From the peak T$_{mb}$ temperatures observed of 5.5, 5.5, 3.6 and 5mK,
respectively,
we deduce filling factors of 0.1, 5.5 10$^{-2}$,  1.6 10$^{-2}$
and 2.2 10$^{-2}$  respectively for the CO(3-2), CO(4-3), CO(6-5)
and 752 GHz H$_2$O emissions, at a given velocity. 
These surface filling factors correspond to a total source size of 
1.5" = 7.8 kpc, since the angular size distance is $D_A$= 1.08 Gpc
(while the luminosity distance is $D_L$ = 11.7 Gpc, with $H_0$ = 75 km/s/Mpc
and $q_0$ = 0.5). This is however only the apparent magnified size
of the source, which could be intrinsically lower (i.e. $\sim$ 1kpc).
Fortunately,
the estimation of the filling factors do not depend on the magnification
values, as far as these are comparable for all mm lines. The relative
filling factor of the dense H$_2$O emitting cores with respect to the
more extended CO(3-2) emitting region is 0.22. This is very large for
emission on kpc scales. By comparison, the size of the thermal HDO emitting
core in Orion is 10", even smaller than the 120" size of the 183 GHz maser 
emission (Jacq et al 1990, Gensheimer et al 1996).

As an order of magnitude, we will assume a comparable surface filling factor 
ratio (0.22) between the 183 GHz H$_2$O and CO(1-0) emission hot core; in Orion,
the hot core CO(1-0) emission (T$_b$(max) $\sim$ 200K) has an extension of the 
order of 12" (Masson et al 1984), even smaller than that of the 183 GHz H$_2$O 
maser emission. With an average brightness temperature over 0.1pc of 
1000 K for the maser and 80K for CO, as observed in Orion 
(Cernicharo et al. 1994), we expect to detect an antenna temperature for
H$_2$O seven times higher than for CO, taking into account the different 
beams.
 This is obviously not observed: on the contrary, from the upper limits
displayed in Table 1, we find at 3$\sigma$ that the signals are 10 and
20 times lower than expected for Mrk1014 and VIIZw244 respectively.

  These low upper limits can be interpreted in two ways:
either the number of
 hot cores of the Orion type in Mrk1014 and
VIIZw244 is an order of magnitude smaller than in the starburst F10214+4724
or the physical conditions are not appropriate to pump efficiently the 
183 GHz line to become as bright as 1000 K.
Another possibility is that the relative H$_2$O/CO filling factor in 
F10214+4724 is overestimated, because of different magnification
factors due to the gravitational lens: i.e. the locations and sizes of the 
H$_2$O and CO dense cores could be different. 
This is not likely however, since the CO magnification factor
is already of the order of 10, and therefore the H$_2$O 
gravitational magnification
would have to be 200, even larger than the optical value.

\begin{figure}
\psfig{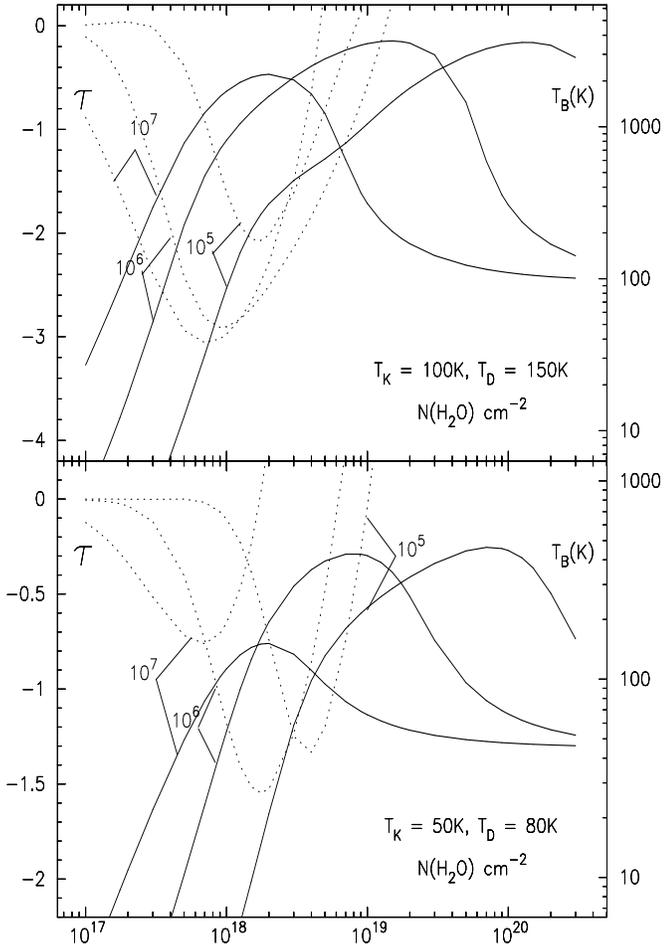}
\caption[]{ 
Results from the LVG radiative transfer model for the
3$_{13}$ -- 2$_{20}$ 183 GHz line of para-water for three different densities
(n(H$_2$)= 10$^5$,10$^6$ and 10$^7$ cm$^{-3}$) and two gas temperatures 
(50 and 100K). The dust temperature is respectively T$_D$= 80 and 150K.
The adopted line width is 20km/s. 
The expected brightness temperatures (solid lines) can 
be read on the right scale, and the optical depth (dotted lines) 
on the left scale.}
\label{maser_fig3}
\end{figure}

\begin{figure}
\psfig{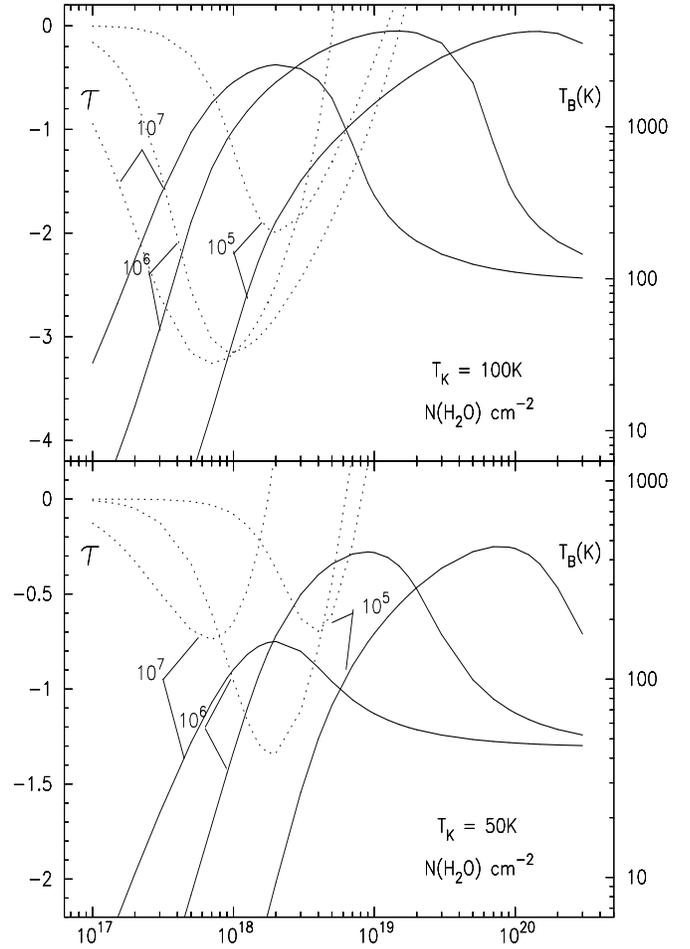} 
\caption[]{ Same as Fig. 3 without taking into account the dust grains}
\label{maser_fig4}
\end{figure}

\subsection { Water vapor excitation }

It was shown that rotational transitions in the ground vibrational state
of water is mainly excited by
collisions with H$_2$  (Deguchi and Nguyen-Q-Rieu, 1990). At a kinetic 
temperature of 100 K, the critical density
to excite the 183 GHz line is  $\sim$ 5 10$^4$ cm$^{-3}$.   
The 183 GHz transition is the lowest level masing transition 
(3$_{13}$ -- 2$_{20}$ of para-H$_2$O). The energy level of the upper state of 
this transition (200 K) is much lower than that of the well-known microwave  
(6$_{16}$ -- 5$_{23}$) maser 
transition (640 K) at 22 GHz.  This explains why the 183 GHz emission region 
in Orion with a peak main beam brightness temperature of 2000 K is so extended,
 about one arcmin in size (0.2 pc), as shown by Cernicharo et al (1994). 
    
In order to investigate the physical conditions in warm and dense cores,
typical of the ULIGs starburst galaxies,
we use a simple Large Velocity Gradient (LVG) model of radiative transfer in
a spherical geometry. The collisional rates between H$_2$ and H$_2$O are 
derived from those of He-H$_2$O calculated by Green et al. (1993) 
by multiplying the latter rates 
by a factor of 1.35. The 25 lowest rotational levels 
in the ground vibrational state are considered in the model.
 An adopted line width of 20 km/s corresponds to the line
width of the 183 GHz H$_2$O maser emission observed in Orion.  

Figure \ref{maser_fig3} shows the optical depth and the 
brightness temperature of
the  183 GHz line as a function of the H$_2$O column density for a 
gas density n(H$_2$) of 10$^5$- 10$^7$
cm$^{-3}$ and for T$_K$ between 50 K and 100 K. The line is inverted in a 
wide range of physical conditions. 
We include the 2.7K cosmic background and the radiation due to dust
 embedded in the molecular clouds. The absorption efficiency of dust 
grains varies as Q$_{abs}(\lambda)$ = Q$_{abs}(80\mu) 
(\lambda/80\mu)^{-p}$, with Q$_{abs}(80\mu)\approx$ 10$^{-4}$- 
10$^{-3}$ and $p\sim$ 1.5-3.5 (Guibert et al 1978). The dilution factor 
at 80$\mu$ is $W= A_v  Q_{abs}(80\mu) $, where $A_v$ is the visual 
extinction. Figure \ref{maser_fig3} corresponds to
$W(80\mu)$ = 5  10$^{-3}$, $p$ = 1.5 for different 
values of the grain temperature T$_D$. The 183GHz line 
is inverted as long as N(H$_2$O) $<10^{19}$ cm$^{-2}$. Beyond this 
limit the maser is quenched. For comparison, figure \ref{maser_fig4} 
shows the results in the absence of dust grains. The pumping of the 
masers is essentially due to collisions. However, radiation by dust can 
contribute to the maser pumping when the gas temperature is low 
(T$_k< 50$K). In the absence of dust radiation, our results are similar 
to those obtained by Cernicharo et al (1994).

The upper limits 
obtained for the H$_2$O line intensities in Mrk1014 and VIIZw244 can
be reproduced only if the temperature is much below 50K (but the average dust
temperature is already 49 and 33K in these two objects respectively).
Also the column density, or the volumic density, could be so high that the 
maser is quenched. Quenching occurs at N(H$_2$O) = 7-8 10$^{18}$ cm$^{-2}$,
for n(H$_2$) = 10$^5$-10$^6$ cm$^{-3}$, and at N(H$_2$O)= 2 10$^{18}$ 
cm$^{-2}$, for n(H$_2$) = 10$^7$ cm$^{-3}$, at T$_k$=50K. Since the H$_2$O
abundance is expected between 10$^{-4}$ and 10$^{-5}$, the corresponding
H$_2$ column densities are very high, of the order of 10$^{24}$ cm$^{-2}$
and more. Other, less likely, explanations for our negative results
could be a very low H$_2$O abundance, or extremely high densities
(larger than 10$^7$ cm$^{-3}$). 
  
\begin{figure}
\psfig{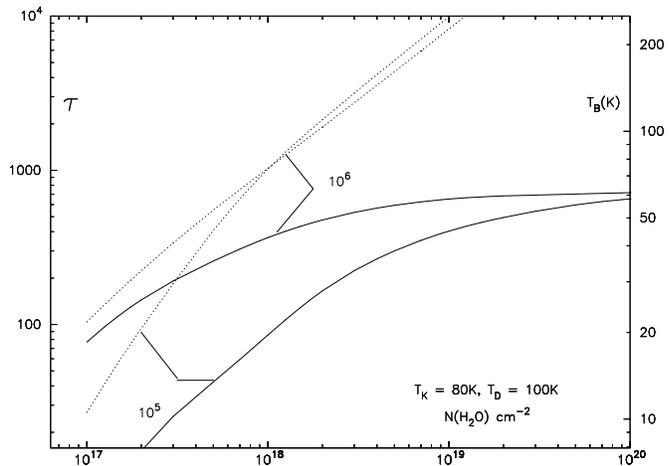} 
\caption[]{ Same as figure 3, but for the 
2$_{11}$ -- 2$_{02}$ 752 GHz line of para-water, observed in F10214+4724 by
Encrenaz et al (1993). The assumed average kinetic temperature for this source
is 80K, and the dust temperature 100K. However, the influence of dust
is not significant. }
\label{maser_fig5}
\end{figure}

We have also tried to reproduce the signal tentatively detected for 
the 2$_{11}$ -- 2$_{02}$ line at 752 GHz in 
IRAS F10214+4724 by Encrenaz et al (1993). The source filling factor
is 2.2 10$^{-2}$, for a rest-frame brightness temperature of 80K. 
Our model calculations with n(H$_2$) = 10$^6$ cm$^{-3}$, T$_K$ = 80 K 
 shows that the line is thermalized to $\sim$60 K (Fig. \ref{maser_fig5}). 
 The energy of the upper level of this transition is only at
137 K. 
The line becomes saturated at high 
water column densities (N(H$_2$O) $>10^{19}$ cm$^{-2}$). The
 excitation by radiation from dust is not significant.

Encrenaz et al. also searched for the 
 4$_{14}$ -- 3$_{21}$ line at 380 GHz in this galaxy without success. 
With the same physical conditions as above, our calculations indicate that
 the 380 GHz line is weak and becomes detectable only with a water column
density higher than 5 10$^{18}$ cm$^{-2}$. This means that the highest
column densities have not a large covering factor in
F10214+4724.

\section {Conclusion}
We have not detected the H$_2$O 183 GHz masing line in two ultraluminous
starburst galaxies, and the 3$\sigma$ upper limits are an order of magnitude
lower than the expected signals. Predictions were based on the
high surface filling factor of warm and dense molecular gas, similar to
the Orion star-forming core, where 183 GHz emission has been detected
with a very high brightness temperature on very extended scales
(Cernicharo et al 1994). The fact hat most of the gas of ultraluminous starburst
should be in warm and dense cores is suggested by the observations in 
these objects of high H$_2$ column densities, large volumic densities,
and high dust temperatures, averaged over one kpc scale.
  This result could be explained if the water column density are so
high as to quench the maser excitation.

Observations of water in distant starburst galaxies with ISO do not seem
promising. For example, the 2$_{12}$ -- 1$_{01}$ line at 179.5 
$\mu$m in Mrk1014 can be estimated to be $\sim$ 3 10$^{-15}$ erg s$^{-1}$
cm$^{-2}$. This predicted flux is $\sim$ 3 magnitudes below the 
detection limit of the Long Wavelength Spectrometer of ISO. The 179 $\mu$m 
line could be detected in absorption however, in front of
a strong enough background continuum source (cf Cernicharo et al 1997).

\bigskip

\begin{acknowledgements}    
We are grateful to Nicole Jacquinet Husson and Claude Camy-Peyret for 
providing us molecular data on H$_2$O and for their advice. Our thanks
go also to J. Cernicharo, the referee, whose comments helped to improve the
paper.
\end{acknowledgements}

\end{document}